\documentclass[smallextended]{svjour3}

\usepackage[colorlinks,allcolors=blue,unicode]{hyperref}

\usepackage{amssymb}
\usepackage{amsxtra}
\usepackage[final]{graphicx}
\usepackage{cite}
\usepackage[utf8]{inputenc}
\usepackage[mathscr]{eucal}
\usepackage[english]{babel}

\hypersetup{
    pdftitle = {Heat transfer in a one-dimensional harmonic crystal},
}

\makeatletter

\@addtoreset{equation}{section}
\makeatother
\journalname{CMAT}
\begin{document}
\selectlanguage{english}
\title{Heat transfer in a one-dimensional harmonic crystal in a viscous environment
subjected to an external heat supply
\thanks{This work is supported by 
Russian Foundation for Basic Research (grant No.~16-29-15121)}
}
\titlerunning{Heat transfer in a one-dimensional harmonic crystal} 
% if too long for running head
\author{S.N.~Gavrilov \and A.M.~Krivtsov \and D.V.~Tsvetkov}
\institute{
S.N.~Gavrilov \at
Institute for Problems in Mechanical Engineering RAS, V.O., Bolshoy pr.~61,
St.~Petersburg, 199178, Russia \\
\email{serge@pdmi.ras.ru}           
\and
S.N.~Gavrilov \at
Peter the Great St.~Petersburg Polytechnic University,
Polytechnicheskaya str.~29, St.Petersburg, 195251, Russia
\and
A.M.~Krivtsov \at
Institute for Problems in Mechanical Engineering RAS, V.O., Bolshoy pr.~61,
St.~Petersburg, 199178, Russia \\
\email{akrivtsov@bk.ru}           
\and
A.M.~Krivtsov \at
Peter the Great St.~Petersburg Polytechnic University,
Polytechnicheskaya str.~29, St.Petersburg, 195251, Russia
\and
D.V.~Tsvetkov \at
Peter the Great St.~Petersburg Polytechnic University,
Polytechnicheskaya str.~29, St.Petersburg, 195251, Russia\\
\email{DVTsvetkov@ya.ru}
}

%\ead{\mailto{serge@pdmi.ras.ru}, \mailto{akrivtsov@bk.ru}, \mailto{DVTsvetkov@ya.ru}}

\selectlanguage{english}
\maketitle
\begin{abstract}	
We consider unsteady heat transfer in a one-dimensional harmonic crystal
surrounded by a
viscous environment and subjected to an external heat supply. The basic equations for
the crystal particles are stated in the form of a system of stochastic
differential equations. We perform a continualization procedure and
derive an infinite set of linear partial differential equations for covariance variables.  An exact
analytic solution describing unsteady ballistic heat transfer in the crystal
is obtained.  It is shown that the stationary spatial profile of the kinetic
temperature caused by a point source of heat supply of constant
intensity is described by the Macdonald function  of zero order.  A
comparison with the results obtained in the framework of the classical heat
equation is presented.  We expect that the results obtained in the paper can be
verified by experiments with laser excitation of low-dimensional
nanostructures.


\end{abstract}

%\pacs{44.10.+i, 05.40.–a, 05.60.–k, 05.70.Ln}
%\input gitdef

\def\F#1#2#3{#1_{{\mathscr F_{#3}^#2}}}
\def\A{\hat a}
\def\varOmega{\hat\omega}
\def\LA{{\mathcal L}^{\mathscr A}}
\def\LS{{\mathcal L}^{\mathscr S}}

\def\be#1{\begin{equation}\label{#1}}
\def\ee{\end{equation}}
\newcommand {\ba}[2]{\be{#1}\begin{array}{#2}}
\newcommand {\ea}{\end{array} \ee}
\def\eq#1{(\ref{#1})}
\newcommand {\hence}{\qquad\Longrightarrow\qquad}
\newcommand {\eqv}  {\qquad\Longleftrightarrow\qquad}
\newcommand{\q}{\,,\quad}
\newcommand{\qq}{\,,\qquad}
\newcommand{\qqp}{\,;~~\qquad}
\renewcommand{\=}{\stackrel{\mbox{\scriptsize def}}{=}}
\let\TS=\textstyle
\let\DS=\displaystyle
\def\({\left(}
\def\){\right)}
\def\av#1{\langle{#1}\rangle}
\def\Av#1{\left\langle{#1}\right\rangle}
\let\w = \omega
\def\kB{k_{\!B}}
\def\k{k}
\def\pti#1{#1}
\def\fig#1{Рис.~\ref{fig-#1}}
\def\figlabel#1{\label{fig-#1}}

\def\dx{\partial_x}
\def\dt{\partial_t}
\def\W{\varOmega}
\def\LL{{\mathfrak L}}
\def\D{{\Delta}}
\def\SS{{\Sigma}}
\let\de = \delta
\let\la = \lambda
\def\L{{\mathcal L}}
\def\M{{\cal M}}
\def\R{{\cal R}}
\def\k{\kappa}
\def\d{{\cal D}}
\let\ov=\overline
\let\ka=\kappa
\let\si=\sigma
\let\ro=\varrho
\def\N{{\cal N}}
\let\al=\alpha
\def\TODO#1{\marginparwidth=15mm
\marginpar{\hrule\strut\vphantom{p}\scriptsize #1\strut\vphantom{p}\hrule}}

\def\DANGER#1{%
\marginpar{\ \ \includegraphics[scale=0.1]{./fig/danger.eps}}{\it #1}%
}
\def\DDANGER#1{%
\marginpar{\includegraphics[scale=0.1]{./fig/danger.eps}\includegraphics[scale=0.1]{./fig/danger.eps}}{\it
#1} }
\def\NOTDANGER#1{#1}
\def\NOTDDANGER#1{#1}
\def\sign{\operatorname{sign}}

\def\qy{{y}}

\def\MYdef{\mathrel{\stackrel{\rm def}=}}
\def\chio{\bar\chi_0}
\def\chil{\bar\chi_1}
\def\erfc{\operatorname{erfc}}

\def\NEW#1{{\color{black}#1}}
\def\ANEW#1{{\color{black}#1}}
\def\LL{\NEW{\omega_0^2\L}}

\def\pd#1#2{\frac{\partial #1}{\partial #2}}
\def\pdd#1#2{\frac{\partial^2 #1}{\partial^2 #2}}
\def\I{\mathrm i}

\section{Introduction}
%\TODO{We need to cite 
%\cite{mielke2006macroscopic,harris2008energy,lukkarinen2016harmonic} et al.
%somehow.} 

An understanding of heat transfer at the microlevel is essential to obtain a
link between the microscopic and the macroscopic descriptions of solids. As
far as the macroscopic level is concerned, Fourier's law of heat
conduction is widely and successfully used to describe heat transfer
processes.
However, it is well known that for one-dimensional crystals substantial deviations from Fourier's law are observed 
\cite{rieder1967properties,%
bonetto2000mathematical,%
lepri2003thermal,%
0295-5075-43-3-271,%
dhar2008heat}.
\ANEW{
Extensive investigations over the last decades were devoted to resolving these
anomalies, many of the recent developments in this area are reviewed in book
\cite{Lepri2016thermal}.
One of the possible solutions is to use special laws of particle
interactions
\cite{casati1984one,%
aoki2000bulk,%
gendelman2014normal,%
savin2014thermal,gendelman2016heat},
in particular, systems on a nonlinear elastic support with no momentum
conservation~\cite{aoki2000bulk}, or systems possessing the possibility of bond break
\cite{gendelman2014normal,savin2014thermal,gendelman2016heat}. Such systems
under certain conditions demonstrate normal heat conductivity even in one
dimension.  As it is shown in \cite{spohn2016fluctuating}, one-dimensional
case is still very specific, therefore, another way to avoid anomalies is to
use sufficiently complex structures and increase the system dimensionality
\cite{bonetto2004fourier,le2008molecular}.  However, recent experimental
observations demonstrate that Fourier's law is indeed violated in
low-dimensional nanostructures
\cite{chang2008breakdown,%
xu2014length,%
hsiao2015micron,%
cahill2003nanoscale,%
liu2012anomalous,%
lepri2016thermal8}, where the ballistic heat transfer is realized.}
%%Since the pioneering work by Rieder, Lebowitz, and Lieb 
%%\cite{rieder1967properties},
%%it is well known that at the microscopic level 
%%%analytical and numerical investigations show 
%%substantial deviations from Fourier’s law are observed
%%\cite{%
%%bonetto2000mathematical,
%%lepri2003thermal,
%%0295-5075-43-3-271,
%%dhar2008heat}.
%These inadequacies can be,
%in principle, addressed by using special laws of particle
%interactions 
%\cite{casati1984one,%
%aoki2000bulk,%
%gendelman2014normal,%
%savin2014thermal,gendelman2016heat}
%or sufficiently complex structures 
%\cite{bonetto2004fourier,le2008molecular}, 
%\NEW{as well as increasing the problem dimension (in \cite{spohn2016fluctuating} it is
%explained why one-dimensional case is very special)}.  
%Recent experimental data, however, demonstrate that Fourier’s
%law is indeed violated in low-dimensional nanostructures 
%\cite{chang2008breakdown,%
%xu2014length,%
%hsiao2015micron,%
%cahill2003nanoscale,%
%liu2012anomalous,%
%lepri2016thermal8}, where ballistic heat conduction is observed.
This fact is in agreement with the phonon theory
\cite{peierls1955quantum,ziman1960electrons},
which relates the heat conductivity with the phonon mean free path.
At the macroscale, the phonon mean free path is a small quantity in comparison with
the characteristic size of the system, but this is not true for microscale and
nanoscale systems \cite{hsiao2013observation}. 
This motivates the interest in the simplest lattice models, in
particular, in harmonic one-dimensional crystals (chains), where these anomalies are most
prominent \cite{kannan2012nonequilibrium,dhar2015heat}.
Problems of this type were previously addressed mainly in the context of
steady-state heat conduction 
\cite{%
bonetto2000mathematical,%
lepri2003thermal,%
0295-5075-43-3-271,%
dhar2008heat,%
rieder1967properties,%
allen1969energy,%
nakazawa1970lattice,%
lee2005heat,%
kundu2010heat,%
lepri2016thermal2,%
bernardin2012harmonic,%
freitas2014analytic,%
freitas2014erratum,%
hoover2013hamiltonian,%
lukkarinen2016harmonic},
%as well as in the context of 
unsteady conduction regimes came into the focus in
\cite{le2008molecular,%
gendelman2012nonstationary,%
%gusev2012wave,
tsai1976molecular,%
ladd1986lattice,%
volz1996transient,%
daly2002molecular,%
gendelman2010nonstationary,%
babenkov2016energy,krivtsov2015heat,krivtsov2014energy}.

Simple lattice models can be used for the analytical investigation of the
thermomechanical processes in solids at the microscale 
\cite{hoover2015simulation,daly2002molecular,krivtsov2003nonlinear,berinskii2016elastic,kuzkin2016lattice},
%%[Hoover 2015, Daly 2002, Krivtsov 2003 CSF, Berinskii 2016, Kuzkin 2016
%%PhylMag],
and, in particular, in the carbon nanostructures
\cite{berinskii2015linear,berinskii2016hyperboloid}.
%[Berinsky \& Krivtsov 2015, Berinsky \& Krivtsov 2016].
One-dimensional systems
due to their simplicity can be used to obtain analytical solutions in a closed
form  without loss of generality
\cite{krivtsov2003nonlinear,%
dhar2008heat,%
Gavrilov-Shishkina-MMS,%
gendelman2010nonstationary},
%[Krivtsov 2003 CSF, Dhar 2008, Gavrilov in a top-rated journals, Gendelman
%2010],
or to get the asymptotic description of
non-stationary processes in media with complex structure
\cite{Gavrilov_SN--ZAMM87N2-2007-p117_127,%
Gavrilov_SN-Shishkina_EV--CMT22-2010-p299_316,%
Gavrilov-Shishkina-ZAMM,%
Gavrilov-Shishkina-MMS,%
shishkina2017stiff,%
Gavrilov-JSV-2012}.
In previous studies
\cite{krivtsov2014energy,krivtsov2015heat,krivtsov2017heat},
%[Krivtsov 2014 DAN, Krivtsov 2015 DAN], 
a new approach was suggested which allows one to solve
analytically non-stationary thermal problems for a one-dimensional harmonic
crystal --- an infinite ordered chain of identical material particles,
interacting via linear (harmonic) forces. In particular, a heat transfer
equation was obtained that differ from the extended heat transfer equations
suggested earlier \cite{chandrasekharalah1986thermoelasticity,tzou2014macro,cattaneo1958forme,vernotte1958paradoxes};
%[12, 58]; 
however, it is in an excellent agreement with
molecular dynamics simulations and previous analytical estimates 
\cite{gendelman2012nonstationary}. 
%[25]. 
Later
this approach was generalized to a number of systems, namely, to a one-dimensional
crystal on an elastic substrate 
\cite{babenkov2016energy},
%[Babenkov 2016], 
and to two and three-dimensional
harmonic crystals \cite{kuzkin2017analytical,Kuzkin-Krivtsov-accepted}.
%[Kuzkin \& Krivtsov 2017].
In the most of above mentioned papers 
\cite{krivtsov2014energy,krivtsov2015heat,babenkov2016energy,kuzkin2017analytical,Kuzkin-Krivtsov-accepted}
%[Krivtsov 2014 DAN, Krivtsov 2015 DAN, Babenkov 2016, Kuzkin \& Krivtsov 2017] 
only
isolated systems were considered.
%
%%
%%In previous studies %by A.M.~Krivtsov and his co-authors
%%\cite{krivtsov2014energy,krivtsov2015heat},
%%a new approach was suggested which allows one to derive macroscopic heat
%%conduction equations for a one-dimensional harmonic crystal
%%(an infinite ordered chain of identical interacting material particles).
%%The law for interparticle forces was assumed to be linear (harmonic). 
%%The obtained
%%
%%equations differ substantially from the heat transfer
%%equations suggested earlier \cite{chandrasekharalah1986thermoelasticity, tzou2014macro};
%%however, they are in an excellent agreement with molecular
%%dynamics simulations and previous analytical estimates
%%\cite{gendelman2012nonstationary}. 
%%This approach can be easily generalized in a straightforward way to a wide class of
%%problems. 
%%Note that in
%%\cite{babenkov2016energy,krivtsov2015heat,krivtsov2014energy,krivtsov2015unsteady,krivtsov2016ioffe}
%%isolated systems were considered. 
The motivation for this paper is to
consider a system that can exchange energy
with its surroundings.
Therefore, now we assume that the crystal is surrounded by a viscous
environment (a gas or a liquid) which causes an additional dissipative term in
the equations of stochastic dynamics for the particles. Additionally, we take into
account sources of heat supply. This is a more realistic model, and
thus
we expect that the theoretical results obtained in the paper can be verified by
experiments with laser excitation of low-dimensional nanostructures
\cite{lepri2016thermal8,liu2012anomalous,cahill2003nanoscale,indeitsev2017two}.
%\NEW{Note that such a setup is similar to the one describing the interaction of
%a crystal with a heat bath. In the latter case the ratio between the noise
%amplitude and dissipation could be related with the thermal bath temperature 
%(see, e.g.~\cite{maes2014second}). However, in this paper we understand the notion
%of the temperature only as the "kinetic temperature", and the stochastic term is related
%with the heat supply (see Section~\ref{Sec-slow})}.

The paper is organized as follows. In Section~\ref{Sec-formulation}, we
consider the formulation of the problem. In Section~\ref{SSec-notation},
some general notation is introduced. In Section~\ref{SSec-stochastic}, we state the basic
equations for the crystal particles in the form of a system of 
stochastic differential equations. In Section~\ref{SSec-covariance}, we
introduce and deal with infinite set of covariance variables. These  
are the mutual covariances of the particle velocities and the
displacements for all pairs of particles. We use the It\^o lemma to derive
(see Appendix~\ref{App-cov}) an infinite deterministic system of ordinary
differential equations which follows from the equations of stochastic dynamics.
This system can be transformed into an infinite system of differential-difference
equations involving only the covariances for \NEW{the particle velocities}.
In Section~\ref{Sec-cont}, we introduce a continuous spatial variable and write
the finite difference operators involved in the equation for
covariances as compositions of finite difference operators and
operators of differentiation.
To do this, we use some identities of the calculus of finite differences (see
Appendix~\ref{AppB}).
In Section~\ref{Sec-slow}, we perform an asymptotic uncoupling
of the equation for covariances. Provided that the introduced continuous spatial
variable can characterize the behavior of the crystal, one can
distinguish between slow motions, which are related to the heat propagation, and
vanishing fast motions
\cite{krivtsov2014energy,babenkov2016energy}, which are not considered in the paper. Slow motions
can be described by a coupled infinite system of second-order hyperbolic partial
differential equations  for quantities which we call the non-local
temperatures. The zero-order non-local temperature, which is proportional to
the statistical dispersion of the particle velocities, 
is the classical kinetic temperature.
In Section~\ref{section-solution}, we obtain an expression for the fundamental
solution for the kinetic temperature and solve the non-stationary problem of
the heat propagation from a suddenly applied point source of constant 
intensity. In contrast to the case of a crystal without viscous environment 
(Section~\ref{section-solution-cons}),
in the case of a crystal surrounded by a viscous environment
(Section~\ref{section-solution-diss}) there exists a
steady-state solution describing the kinetic temperature distribution caused
by a constant point source. 
In Section~\ref{Sec-numerics}, we present the
results of the numerical solution of the initial value problem for the system
of stochastic differential equations and compare them with the obtained analytical
solution.
In Section~\ref{sec-comparison}, 
we compare our results with  the classical results obtained in the
framework of the heat equation based on Fourier's law. 
In the conclusion (Section~\ref{Sec-conclusion}), we discuss the basic
results of the paper.

\section{Mathematical formulation}
\label{Sec-formulation}
\subsection{Notation}
\label{SSec-notation}
In the paper, we use the following general notation:
\begin{description}	
\item[$t$]  the time;
\item[$H(\cdot)$]  the Heaviside function;
\item[$\delta(\cdot)$]  the Dirac delta function;
\item[$\langle\cdot\rangle$]  the expected value for a random quantity;
\item[$\de_{pq}$]
is the Kronecker delta ($\de_{pq}= 1$ if  $p=q$, and $\de_{pq} = 0$
otherwise);
\item[$\delta_n$]   $\de_n = 1$ if $n=0$ and $\de_n=0$ otherwise;
\item[$J_0(\cdot)$]  the  Bessel function of the first kind of zero order
\cite{andrewsspecial};
\item[$I_0(\cdot)$]  the modified Bessel function of the first kind of zero order
\cite{andrewsspecial};
\item[$K_0(\cdot)$]  the Macdonald function (the modified Bessel function of the second
kind) of zero order
\cite{andrewsspecial};
\item[$\erfc(\cdot)$]  the complementary error function
\cite{andrewsspecial}.
\end{description}

\subsection{Stochastic crystal dynamics}
\label{SSec-stochastic}

Consider the following system of stochastic ordinary differential equations
\cite{kloeden1999,stepanov2013stochastic}:
\begin{gather}
d v_i = F_i dt + b_i d W_i,
 \qquad
d u_i = v_i dt,
\label{1}
\end{gather}
where
\begin{gather}
F_i=\LL_{i} u_i - \eta v_i,
\label{F_i}
\\
d W_i= \rho_i \sqrt{dt}
\label{Winer},\\
\w_0 \= \sqrt{C/m},
\end{gather}
%\TODO{$\mathscr L_{(i)}$?}%
%\TODO{``Harmonic crystal'' or ``harmonic lattice''?}%
\NEW{Here $i$ is an
arbitrary integer which
describes the position of a particle in the chain; 
the stochastic processes $u_i(t)$ and $v_i(t)$ are the displacement and
the particle velocity, respectively;
$F_i$ is the specific force on the
particle;  
$W_i$ are Wiener processes;
$b_i(t)$ is the intensity of the random external excitation;
$\eta$ is the specific viscosity for the environment;
$C$ is the bond stiffness; $m$ is the mass of a particle;
\NEW{$\L_i$} is the linear finite
difference operator:
\begin{gather}
%\LL_{i} u_i = \w_0^2\L_i u_i,
%\label{LL-L}\\
\L_i u_i=u_{i+1} - 2u_{i} + u_{i-1}.
\label{Li}
\end{gather}
Note that the results of the paper can be generalized for more
complex finite difference operators and related physical systems (e.g. a
crystal on an elastic support, next neighbour interactions etc).}

The normal random variables $\rho_i$ are such that 
\begin{equation}
    \av{\rho_i} = 0, \qquad
    \av{\rho_i\rho_j} = \de_{ij},
    \label{82}
\end{equation}
and they are assumed to be independent of $u_i$ and $v_i$.
%where the angular brackets stand for the expected value. 
The initial conditions are zero: for all $i$,
\begin{equation}
u_i(0)=0,\qquad v_i(0)=0.
\label{ic-stochastic}
\end{equation}

\begin{figure}[htp]
\centering\includegraphics[width=0.6\textwidth]{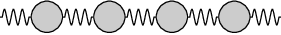}
\caption{A one-dimensional harmonic crystal.}
\label{fig-crystal}
\end{figure}

In the case $b_i\equiv b$, equations \eq{1} are the Langevin equations
\cite{langevin1908theorie,lemons1997paul} for a
one-dimensional harmonic crystal (an ordered chain of identical interacting
material particles, see Fig.~\ref{fig-crystal})
surrounded by a viscous environment (e.g., a gas or a liquid). Assuming that
$b_i$ may depend on $i$, we introduce a natural generalization of the Langevin
equation which allows one to describe the possibility of an external heat
excitation (e.g.,\ laser excitation). 
\NEW{This external excitation is assumed to be localized in space (in
the paper we mostly consider the case of a point heat source) 
and much more intensive than the stochastic influence caused by a non-zero
temperature of the environment. Therefore,
we neglect 
in 
\eqref{1}
the constant stochastic term that does not depend on $i$}.
Note that since $b_i$ do not depend on
$u_j$ and $v_j$ for all $i,j$, it is not necessary to distinguish between the Stratonovich and It\^o
formalism \cite{kloeden1999} in the case of equation \eqref{1}.

\subsection{The dynamics of covariances}
\label{SSec-covariance}

According to \eqref{F_i}, $F_i$ are linear functions of $u_i,\ v_i$. Taking
this fact into account together with \eqref{82} and \eqref{ic-stochastic}, 
we see that for all $t$
\begin{equation}
\av{u_i}=0,\qquad \av{v_i}=0.
\end{equation}
%% \DANGER{In what follows, we assume that expected values for displacements and particle
%% velocities are zero.} 
%\TODO{Is \cite{krivtsov2016ioffe} enough?}
Following \cite{krivtsov2016ioffe}, consider the infinite sets of
covariance variables
\begin{equation}
    \xi_{p,q} \= \av{u_p u_q}, \qquad
    \nu_{p,q} \= \av{u_p v_q}, \qquad
    \ka_{p,q} \= \av{v_p v_q},
\label{4}
\end{equation}
and the \NEW{quantities}
\begin{equation}
\beta_{p,q} \= \de_{pq}\,b_p b_q.
\label{beta-pq}
\end{equation}
In the last equation, we take into account the second equation of \eqref{82}.
Thus \NEW{the variables $\xi_{p,q},\ \nu_{p,q},\ \ka_{p,q}$ and $\beta_{p,q}$} are defined
for any pair of crystal particles.
For simplicity, in what follows we drop the
subscripts $p$ and $q$, i.e., $\xi\=\xi_{p,q}$ etc. By definition, we also put
$\xi^\top\=\xi_{q,p}$ etc. 
Now we differentiate the variables 
\eqref{4} with respect to time
taking into account the equations of motion \eqref{1}. This yields
the following closed system of differential equations for
the covariances (see Appendix~\ref{App-cov}):
%\footnote{\foreignlanguage{russian}{Строго говоря, при воздействии Ланженвена
%коэффициенты $\eta$ и $b$ связаны между собой, и зависимость от времени одной
%из этих величин влечет за собой зависимость от времени другой. Однако,
%зависимость $b$ от времени, в отличие от $\eta$, не сильно усложняет уравнения
%и позволяет сравнивать результаты с полученными ранее формулами, описывающими
%фотонное воздействие на одномерный кристалл \cite{Krivtsov 2014 Stt, Krivtsov
%2016 ICTAM}.}}
\begin{equation}	
\begin{gathered}	
    \dt\xi = \nu + \nu^\top,\\
    \dt\nu + \eta\nu = \LL_q\xi + \ka,\\
    \dt\ka + 2\eta\ka = \LL_p\nu + \LL_q\nu^\top + \beta,
\end{gathered}
\label{5}
\end{equation}
where $\dt$ is the operator of differentiation with respect to time;
\NEW{$\L_p$ and $\L_q$ are the linear difference operators defined by 
\eqref{Li} that act on  
$\xi_{p,q},\ \nu_{p,q},\ \ka_{p,q},\ \beta_{p,q}$
with respect to the first index subscript $p$ and the second one $q$, respectively.}
%\ba{3}c
%    \dot\xi_{pq} = \nu_{pq} + \nu_{qp} \qq
%    \dot\nu_{pq} + \eta_q\nu_{pq} = \L_q\xi_{pq} + \ka_{pq}
%    , \\ [5mm]
%    \dot\ka_{pq} + (\eta_p+\eta_q)\ka_{pq} = \L_p\nu_{pq} + \L_q\nu_{qp} + \beta_{pq}.
%\ea
%
%%\subsection{Однородная среда}
%
%В предположении, что величина $\eta_{p}$ не зависит от $p$ 
%\DANGER{(кристалл в однородной
%внешней среде)} и не зависит от времени  система уравнений \eq{3} приобретает вид
%%где $\beta_{pq} = b^2\de_{pq}$.
% Introducing the notation
% \begin{gather}
%     \xi \= \xi_{pq} \qq
%     \nu_1 \= \nu_{pq} \qq
%     \nu_2 \= \nu_{qp} \qq
%     \ka \= \ka_{pq} \qq 
%     \beta \= \beta_{pq} \qq \\
%     \L_1 \= \L_{p} \qq
%     \L_2 \= \L_{q},
% \end{gather}
% one can rewrite Eq.~\eq{5} in the form of
% \begin{equation}
% \begin{aligned}
% &\dot\xi_{pq} = 2\nu_{pq} + \nu_{qp},
% &&\dot\nu_{pq} + \eta\nu_{pq} = \L_q\xi_{pq} + \ka_{pq}, \\
% &\dot\ka + 2\eta\ka = \L_1\nu_1 + \L_2\nu_2 + \beta \qq
% &&\dot\nu_2 + \eta\nu_2 = \L_1\xi + \ka .
% \end{aligned}
% \label{7}
% \end{equation}	
%где $\beta = b^2\de_{pq}$.
%\subsection{Переобозначение}
%\subsection{Переход к операторам $\L$ и $\M$}
Now we introduce the symmetric and antisymmetric difference operators
\begin{equation}
%2\LL^{\mathrm S} \= \LL_p+\LL_q,\qquad
%2\LL^{\mathrm A} \= \LL_p-\LL_q,\\
% \qquad
% 2\LL^{\mathrm A\top} \= \LL_q-\LL_p,
2\L^{\mathrm S} \= \L_p+\L_q,\qquad
2\L^{\mathrm A} \= \L_p-\L_q,
\end{equation}
and the symmetric and antisymmetric parts of the variable $\nu$:
\begin{equation}
    2\nu^{\mathrm S} \= \nu+\nu^\top, \qquad
    2\nu^{\mathrm A} \= \nu-\nu^\top.
%     \qquad
%     2\nu^{\mathrm A\top} \= \nu^\top-\nu.
\label{9}
\end{equation}
Note that $\xi$ and $\kappa$ are symmetric variables.
Now equations \eq{5} can be rewritten as follows: 
%\begin{equation}
%\begin{gathered}	
%    \dot\xi = 2\nu \qq
%    \dot\nu + \eta\nu = \L\xi + \ka \qq
%    \dot\mu + \eta\mu = -\M\xi,\\
%    \dot\ka + 2\eta\ka = 2(\L\nu + \M\mu) + \beta.
%\end{gathered}
%\label{10}
%\end{equation}
%
%\section{Уравнения 2-го и 4-го порядка}
%
%\subsection{Вывод уравнений}
%
%Запишем уравнения \eq{10} в операторной форме
\begin{gather}
\dt\xi = 2\nu^{\mathrm S},
\qquad(\dt + 2\eta)\ka =  2\LL^{\mathrm S}\nu^{\mathrm S} 
+ 2\LL^{\mathrm A}\nu^{\mathrm A} + \beta,
\label{11}
\\
(\dt + \eta)\nu^{\mathrm A} = -\LL^{\mathrm A}\xi,
\qquad(\dt + \eta)\nu^{\mathrm S} = \LL^{\mathrm S}\xi + \ka.
\label{11-post}
\end{gather}
Applying the operator $\dt + \eta$ to Eqs.~\eqref{11} and
substituting expressions
\eqref{11-post}
yields a closed system
of two equations of second order in time:
\begin{gather}	
    \dt(\dt + \eta)\xi = 2(\LL^{\mathrm S}\xi + \ka),\label{12--1}\\
    (\dt + \eta)(\dt + 2\eta)\ka 
    = 2\big((\LL^{\mathrm S})^2-(\LL^{\mathrm  A})^2\big)\xi 
    + 2\LL^{\mathrm S}\ka + (\dt + \eta)\beta.
\label{12--2}
\end{gather}
\NEW{We can express $\ka$ in terms of $\xi$ using Eq.~\eqref{12--1}:
\begin{equation}
    \ka  = \frac12(\dt^2 + \eta\dt - 2\LL^{\mathrm S})\xi,
\label{29}
\end{equation}
and substitute the result into Eq.~\eqref{12--2}. This yields 
\begin{multline}
    \frac12\Big(\big((\dt + \eta)(\dt + 2\eta)-
     2\LL^{\mathrm S}\big)(\dt^2 + \eta\dt - 2\LL^{\mathrm S})
     - 4\big((\LL^{\mathrm S})^2-(\LL^{\mathrm  A})^2\Big)\xi 
    \\= (\dt + \eta)\beta.
\label{podrob}
\end{multline}
Simplifying the left-hand side of 
Eq.~\eqref{podrob}
results in
an equation of
fourth order in time for $\xi$:
%%% %%%     \iffalse
%%% %%%     \ba{13a}{c}
%%% %%%     %    \frac14\((\dt + \eta)(\dt + 2\eta)-2\L\)\(\dt(\dt + \eta) - 2\L\)\xi = (\L^2+\M^2)\xi.
%%% %%%         \frac14\(\dt^2 + \eta\dt - 2\L\)\(\dt^2 + 3\eta\dt + 2\eta^2-2\L\)\xi = (\L^2-\M^2)\xi + \frac12\eta\beta.
%%% %%%     \ea
%%% %%%     или
%%% %%%     \ba{13}{c}
%%% %%%         \((\dt^2 + \eta\dt)(\dt^2 + 3\eta\dt + 2\eta^2) - 4(\dt^2 + 2\eta\dt + \eta^2)\L + 4\M^2\vphantom{\Bigl|}\)\xi = 2(\dt + \eta)\beta
%%% %%%     \ea
%%% %%%     или
%%% %%%     \ba{13a}{c}
%%% %%%         \(\dt(\dt + \eta)^2(\dt + 2\eta) - 4(\dt + \eta)^2\L + 4\M^2\vphantom{\Bigl|}\)\xi = 2(\dt + \eta)\beta
%%% %%%     \ea
%%% %%%     или
%%% %%%     \fi
\begin{equation}
\big((\dt + \eta)^2(\dt^2 + 2\eta\dt - 4\LL^{\mathrm S}) 
+ 4(\LL^{\mathrm A})^2\big)\xi =
    2(\dt + \eta)\beta.
\label{13}
\end{equation}
Now we apply the operator 
$\frac12(\dt^2 + \eta\dt - 2\LL^{\mathrm S})$
to Eq.~\eq{13}. Taking into account~\eq{29},
this yields a fourth-order equation for the covariances of the particle
velocities~$\ka$:}%
%\ba{30}{c}
%    \((\dt^2 + \eta\dt)(\dt^2 + 3\eta\dt + 2\eta^2) - 4(\dt^2 + 2\eta\dt + \eta^2)\L + 4\M^2\vphantom{\Bigl|}\)\!\ka =
%%    \\ [4mm] =
%    (\dt\!+\!\eta)(\dt^2 + \eta\dt - 2\L)\beta.
%\ea
\begin{equation}
\((\dt + \eta)^2(\dt^2 + 2\eta\dt - 4
%\omega_0^2
\LL^{\mathrm S}) 
+ 4
%\omega_0^4
(\LL^{\mathrm A})^2\)\ka =
(\dt+\eta)(\dt^2 + \eta\dt - 2
%\omega_0^2
\LL^{\mathrm S})\beta.
\label{30}
\end{equation}
%\TODO{Do we need the equation for $\nu$?}
In what follows, we deal with Eq.~\eqref{30}. According to 
Eqs.~\eqref{ic-stochastic}, \eqref{4}, we supplement Eq.~\eqref{30}
with zero initial conditions. We state these conditions in the following form, 
which is conventional for distributions (or generalized functions)~\cite{Vladimirov1971}:
\begin{equation}
\kappa\big|_{t<0}\equiv0.
\label{ic<0-pre}
\end{equation}
To take into account non-zero classical initial conditions, one needs to add 
the corresponding singular terms (in the form of a linear combination of 
$\delta(t)$ and its derivatives) to the right-hand sides of the corresponding equations 
\cite{Vladimirov1971}.

\NEW{Let us note that  equation \eqref{30} is
a determenistic equation}. What is also important is that \eqref{30} is a closed
equation. Thus the thermal processes do not depend on any property of the
cumulative distribution functions for the displacements and the particle
velocities other than the covariance variables used above.

\section{Continualization of the finite difference operators}
\label{Sec-cont}

In this section, we use some identities of the calculus of finite differences (see
Appendix~\ref{AppB}).
Following \cite{krivtsov2015heat,krivtsov2016ioffe}, we 
introduce the discrete spatial variable
\begin{equation}
k\= p+q
\label{k-def}
\end{equation}
and the discrete correlational variable
%\TODO{уже введен после \eqref{f12}}
\begin{equation}
     n\=q-p
\label{n}
\end{equation}
instead of discrete variables $p$ and $q$.
We can also formally introduce the continuous spatial variable 
\begin{equation}
x\=\frac{ak}2,
\end{equation}
where $a$ is {the lattice constant} (the distance between neighboring  particles). 
We have
\begin{equation}
q=\frac xa+\frac n2,\qquad p=\frac xa-\frac n2.
\label{pq-nk}
\end{equation}

%Operator $\L_i$ can be represented in the following
%form (see~\eqref{AppB-D2})
%%\TODO{What is half-integer?}
%\begin{gather}
%\L_i=\D_i^2.
%\label{factor-L}
%%% \\
%%% \D_i u_i =u_{i+1/2}-u_{i-1/2}.
%\end{gather}
%Here $\D_i$ is defined by means of the first formula in \eqref{AB3} and
%\eqref{AB1}.
%Now, one has 
%\begin{gather}
%\NEW{
%2(\L^{\mathrm A})=\L_p-\L_q=
%\mathscr D_n \mathscr D_k
%}
%%(\D_p-\D_q)^2(\D_p+\D_q)^2
%%=4 (\D^{\mathrm S})^2(\D^{\mathrm A})^2
%\label{LA}
%,\\
%2\L^{\mathrm S}=\D_p^2+\D_q^2,
%\label{LS}
%\end{gather}
%\NEW{
%}

%\DANGER{One can prove the identities (see Appendix)}

To perform the continualization, we assume that the lattice constant is an
infinitesimal quantity and introduce a dimensionless formal small parameter
$\epsilon$ in the following way:
\begin{gather}
a=\epsilon \hat a,
\label{e-a}
\end{gather}
where $\A=O(1)$. 
To preserve the speed of sound in the crystal
$c\=a\w_0$ as a quantity of order $O(1)$, we additionally assume that
%\TODO{Explain this in more detail}
\begin{gather}
\omega_0=\epsilon^{-1}\varOmega_0,
\label{epsilon}
\end{gather}
where $\varOmega_0=O(1)$. Thus
$c=\hat a\varOmega_0=O(1)$.
The basic assumption that allows one to perform the continualization is
that any \NEW{quantity} $\zeta_{p,q}$ defined by \eqref{4} 
or \eqref{beta-pq}, 
where $p$ and $q$ are defined by 
\eqref{pq-nk},  can be calculated as a value of a smooth function
$\hat\zeta_n(x)$ of the continuous spatial slowly varying coordinate
\begin{equation}
x=\frac{\epsilon\A k}2
\label{x-def}
\end{equation}
and the discrete correlational
%\TODO{$\epsilon \A k$}
variable $n$:
\begin{equation}
\hat\zeta_n(x)\=
\zeta_{k+\tfrac n2, k-\tfrac n2}=
\zeta_{p,q}.
\label{Z_n}
\end{equation}
In accordance with 
\eqref{Z_n},
one has
\begin{equation}
\begin{gathered}	
\L_p\zeta_{p,q}=
\hat\zeta_{n-1}
\left(x+\tfrac a2\right)
-
2\hat\zeta_n(x)
+\hat\zeta_{n+1}
\left(x-\tfrac a2\right)
,\\
\L_q\zeta_{p,q}=
\hat\zeta_{n+1}
\left(x+\tfrac a2\right)
-
2\hat\zeta_n(x)
+\hat\zeta_{n-1}
\left(x-\tfrac a2\right).
\end{gathered}
\end{equation}
Applying the Taylor theorem to these formulas yields
\begin{equation}
\begin{aligned}	
&\L_p\zeta_{p,q}=
\L_n\hat\zeta_{n}
+\tfrac a2\,\partial_x(\hat\zeta_{n-1}-\hat\zeta_{n+1})
\\&\qquad\qquad\qquad+\tfrac {a^2}8\,\partial_x^2(\hat\zeta_{n-1}+\hat\zeta_{n+1})+o(\epsilon^2)
,\\
&\L_q\zeta_{p,q}=
\L_n\hat\zeta_{n}
+\tfrac a2\,\partial_x(\hat\zeta_{n+1}-\hat\zeta_{n-1})
\\&\qquad\qquad\qquad+\tfrac {a^2}8\,\partial_x^2(\hat\zeta_{n+1}+\hat\zeta_{n-1})+o(\epsilon^2).
\end{aligned}
\label{LpLq-cont}
\end{equation}
\ANEW{
An alternative way of continualization can be realized by letting
the number of particles diverge, rather than invoking an increasingly
small separation \cite{lepri2010nonequilibrium}.
Despite the algorithmic difference, these approaches lead to the same result.}

Now we perform the continualization of the operators 
$\L^{\mathrm S},\ \L^{\mathrm A}$.
%defined by
%\eqref{LA}, \eqref{LS}.  
%To do this we use formulas \eqref{Dpq} and 
%\eqref{AppC-fin}. Due to \eqref{AppC-fin}
%\begin{equation}
%\D^2_{2k}\zeta_n(\epsilon \A k)=\big(\tfrac{\epsilon\A}2\partial_{x}+o(\epsilon^2)\big)^2
%\zeta_n(\epsilon \A k)=
%\big(\tfrac a2 \,\partial_x \big)^2\zeta_n(x)+o(\epsilon^2),
%\label{continua}
%\end{equation}
%where 
%%$\L_n$ --- оператор $\L_p$, примененный к индексу $n$;
%$\dx$ is the derivative with respect to spatial variable~$x$.
%% Equations 
%% \eqref{Z_n}, 
%% \eqref{continua}
%% together result in 
%% \begin{equation}
%% \D^2_{2k}\zeta\left[k+\tfrac n2, k-\tfrac n2\right]=\tfrac {a^2}4\, \partial_x^2
%% \zeta\left[\tfrac xa+\tfrac n2, \tfrac xa-\tfrac n2\right]+o(\epsilon^4).
%% \end{equation}
%Using formulas \eqref{Dpq} one obtains now 
%\begin{equation}
%\begin{gathered}
%\D_p^2+\D_q^2=2\D_n^2+O(\epsilon^2),\\
%(\D_p+\D_q)^2=\tfrac{a^2}4(\D^2_{n}+4)\partial_x^2+o(\epsilon^2),\\
%%\D_p\D_q=\D^2_{2k}-\D_n^2,\\
%(\D_p-\D_q)^2=4\D^2_{n}+O(\epsilon^2),
%\end{gathered}
%\end{equation}
Using
\eqref{LpLq-cont}, we obtain
%Therefore, due to \eqref{factor-L}, \eqref{LA}, \eqref{LS},
\begin{gather}	
\L^{\mathrm S}
%=\D_n^2+O(\epsilon^2)
=\L_n+O(\epsilon^2),
\label{L-S}
\\
%(\L^{\mathrm A})^2=a^2\D^2_{n}(\D^2_{n}+4)\partial_x^2+o(\epsilon^2).
\L^{\mathrm A}=-\tfrac a2\,\mathscr D_{n}\partial_x+O(\epsilon^2),
\label{LA-pre-pre}
\end{gather}
where 
\begin{equation}
\mathscr D_n f_n=f_{n+1}-f_{n-1}.
\end{equation}
Now we calculate $(\L^{\mathrm A})^2$ using 
\eqref{LA-pre-pre},
\eqref{AppBD2},
\eqref{AppBD2S2}.
This yields
%\DANGER{One has (see Appendix)}
%\begin{gather}
%\D^2_{n}+4=\SS_n^2,\\
%\SS_n u_n \=u_{n+1/2}+u_{n-1/2},\\
%\SS_n^2=-(-1)^n\D^2_n(-1)^n.
%\end{gather}
%Finally, taking into account 
%\eqref{LA-pre}, one has
\begin{gather}
(\L^{\mathrm A})^2=a^2\M+o(\epsilon^2),
\label{LA2}
\\
\M\=-\frac{1}4\L_{n}(-1)^n\L_{n}(-1)^n\partial_x^2.
\label{M02}
\end{gather}

%\section{Asymptotic splitting of the equations}
\section{Slow motions}
\label{Sec-slow}

%\eq{30}
%\ba{52}{c}
%    \((\dt + \eta)^2(\dt^2 + 2\eta\dt - 4\L) + 4\M^2\vphantom{\Bigl|}\)\ka =
%    (\dt\!+\!\eta)(\dt^2 + \eta\dt - 2\L)\beta.
%\ea
Taking into account 
assumption \eqref{epsilon},
Eq.~\eqref{30} can be rewritten in the following form:
\begin{multline}
\bigg((\dt + \eta)^2\big(\epsilon^2(\dt^2 + 2\eta\dt) -
4\varOmega_0^2\L^{\mathrm S}\big) 
+ 4\frac{\varOmega_0^4}{\epsilon^2}(\L^{\mathrm A})^2\bigg)\ka \\=
(\dt+\eta)\big(\epsilon^2(\dt^2 + \eta\dt) - 2\varOmega_0^2\L^{\mathrm S}\big)\beta.
\label{slow-epsilon}
\end{multline}
% 
% At first, consider the case $\eta=0$. Instead of 
% \eqref{slow-epsilon} one has
% \begin{equation}
% \bigg(\dt^2\big(\epsilon^2\dt^2 - 4\varOmega_0^2\L^{\mathrm S}\big) 
% + 4\frac{\varOmega_0^4}{\epsilon^2}(\L^{\mathrm A})^2\bigg)\ka =
% \dt\big(\epsilon^2\dt^2  - 2\varOmega_0^2\L^{\mathrm S}\big)\beta.
% \end{equation}
% The equation obtained is a differential equation whose highest derivative with
% respect to $t$ is multiplied by a small parameter. 
% \begin{equation}
% \epsilon^2\dt^2\ll\varOmega_0^2\L^{\mathrm S}.
% \end{equation}
% Now, taking into account 
% \eqref{e-a}, \eqref{LA2} results in
% the equation for slow motions in the conservative case
% \cite{?}
% \begin{equation}
%     2(\dt^2\L^{\mathrm S} - c^2\M_0^2)\ka = \dt\L^{\mathrm S}\beta,
% \label{55}
% \end{equation}
% 
% 
%Take $\eta>0$, where $\eta=O(1)$. 
Equation~\eqref{slow-epsilon} is a differential equation whose highest
derivative with respect to $t$ is multiplied by a small parameter. Therefore,
one can expect the existence of two types of solutions, namely, solutions slowly varying in time
and fast varying in time \cite{nayfeh2008perturbation}. 
\NEW{The presence of fast and slow motions is a standard property of
statistical systems. Fast motions are oscillations
of temperature caused by equilibration of kinetic and potential energies. Slow
motions are related with macroscopic heat propagation.}

Considering slow
motions, we assume that 
\begin{equation}
\epsilon^2(\dt^2+2\eta\dt)\kappa\ll\varOmega_0^2\L^{\mathrm S}\kappa,
\qquad
\epsilon^2(\dt^2+\eta\dt)\beta\ll\varOmega_0^2\L^{\mathrm S}\beta.
\label{slow-condition}
\end{equation}
Vanishing solutions that characterize fast motions, which do not satisfy 
\eqref{slow-condition}, are not considered in this paper.
In \cite{krivtsov2014energy}, the properties of fast motions are investigated in
the case of the system under consideration without viscous environment ($\eta=0$)
and external heating ($\beta=0$). 
In \cite{babenkov2016energy}, fast motions in a 
one-dimensional harmonic crystal on an elastic substrate are considered (again
under the zero external action condition). 

Now, taking into account 
\eqref{LA2}, we drop the higher order terms and rewrite equation 
\eqref{slow-epsilon} in the form of an equation for slow motions:
% \TODO{\foreignlanguage{russian}{Утверждение о том, что $\dt\sim\eta$ категорически неверно. Верное
% утверждение: $\eta=O(\dt)$; именно оно и используется по существу дела.}}
%
%Воспользуемся заменой
%\ba{53}{c}
%    \ka = e^{-\eta t}\ka_\eta \hence \dt^n\ka = e^{-\eta t}(\dt-\eta)^n\ka_\eta.
%\ea
%Тогда уравнение \eq{52} примет вид
%\ba{54}{c}
%    \(\dt^2(\dt^2 - 2\eta\dt - 4\L) + 4\M^2\vphantom{\Bigl|}\)\ka_\eta =
%    \dt(\dt^2 - \eta\dt - 2\L)\beta_\eta,
%\ea
%где $\beta_\eta = e^{\eta t}\beta$.
%Для рассмотрения \DDANGER{медленных движений 
%будем считать $\dt\sim\eta$ и сохраним слагаемые до $\eta^2$ включительно}.
%Тогда уравнения \eq{53} примут вид
%В результате для $\ka_\eta$ получаем уравнение
%\ba{56}{c}
%    \L\ddot\ka_\eta - \M^2\ka_\eta =
%    \L\dot\beta_\eta.
%\ea
%Таким образом, для $\ka_\eta$ получаем уравнение, полностью соответствующее консервативной задаче, что позволяет воспользоваться ее решением.
%
%Соответствующее уравнение для $\ka$ имеет вид
\begin{equation}
    2\big((\dt + \eta)^2\L^{\mathrm S} - c^2\M\big)\ka =
    (\dt+\eta)\L^{\mathrm S}\beta.
\label{57}
\end{equation}
Applying the operator $(\L^{\mathrm S})^{-1}$ to Eq.~\eqref{57} results in
\begin{equation}
    \ddot\ka +2\eta\dot\ka + \big(\eta^2 - c^2(\L^{\mathrm S})^{-1}\M\big)\ka =
    \frac12(\dot\beta +\eta\beta) + C,
\label{58}
\end{equation}
where $\L^{\mathrm S} C\equiv 0$. 
\NEW{Here and in what follows we use more compact notaton: the overdot means $\partial_t$,
the prime means $\partial_x$.}
Taking into account the initial conditions
in the form of Eq.~\eqref{ic<0-pre}, one can show that 
%\DANGER{Without loss of generality we can put}
$C=0$.

Now we perform the continualization of the equations.
According to Eqs.~\eqref{L-S}, \eqref{LA2}, \eqref{M02},
we have
%~\cite{Krivtsov 2015 L}
\begin{equation}
    (\L^{\mathrm S})^{-1}\M = -\frac14 (-1)^n
    \L_{n}(-1)^n\dx^2+o(\epsilon^2).
\label{39}
\end{equation}
Now we multiply \eq{58} by $(-1)^nm k_B^{-1}$ \NEW{(here $\kB$ is the Boltzmann
constant)}, and rewrite Eq.~\eq{58} in the following form:
\be{40}\TS
\ddot\theta_n + 2\eta\dot\theta_n 
+\eta^2\theta_n
+ \frac{c^2}4  \L_{n}\theta_n'' = (\dot\chi + \eta\chi)\de_n,
\ee
where, according to \eqref{Z_n}, we introduce the following quantities
depending on the continual spatial variable $x$:
\begin{gather}
     \theta_n(x,t) \= (-1)^n mk_B^{-1} \hat\ka_{n}(x,t),
\label{temp-def}     
\\
     \chi(x,t) \= \tfrac12m k_B^{-1} \hat\beta_{0}(x,t).
\label{chi-def}
\end{gather}
We call $\theta_n$ the non-local temperatures
and identify $\chi$ as the heat supply intensity (note that
$\hat\beta_n\equiv0$ for $n\geq1$ due to 
\eqref{beta-pq}).
Also, we identify $T\=\theta_n\big|_{n=0}$ as the kinetic
temperature, since in the framework of the kinetic theory of gases expression
\eqref{temp-def} for $\theta_0$      
coincides with the expression for the temperature of an ideal gas 
consisting of particles with one degree of freedom.
%\TODO{\foreignlanguage{russian}{Вопрос о двойке}}
% \be{41}
% \ee
% $\theta_n$ is the non-local temperatures,
% \TODO{What is the non-local temperatures} 
% \TODO{What is the kinetic temperature} 
%$n\=q-p$ --- корреляционный индекс,
%\TODO{уже введен после \eqref{f12}}
Now we recall the explicit form \eqref{Li} for $\L_n$ and rewrite Eq.~\eqref{40} 
as the infinite system of partial differential equations
\begin{gather}	
     \ddot\theta_n + 2\eta\dot\theta_n  
+\eta^2\theta_n
+\frac{c^2}4  \big(\theta_{n-1} - 2\theta_{n} + \theta_{n+1}\big)''
     = \phi(x,t)\de_n,
\label{maineq}
\\
\phi=\dot\chi + \eta\chi,
\label{phi-chi}
\end{gather}
which describe the heat propagation in the crystal. The particular case of
this equation for the case $\eta=0,\ \chi\equiv0$ was obtained previously in
\cite{krivtsov2015unsteady}.

Note that equations \eqref{maineq} for slow motions involve only the product
$c=\omega_0a$ and do not involve the quantities $\omega_0$ and $a$ separately, 
so they do not involve $\epsilon$. Provided that the initial conditions
also do not involve $\epsilon$, the solution of the corresponding initial
value problem and all its derivatives are quantities of order $O(1)$.  The
rate of vanishing for fast motions depends on $\epsilon$: the smaller
$\epsilon$, the higher the rate. Thus, for sufficiently small $\epsilon$,  exact
solutions of Eq.~\eqref{30}
quickly transform into slow motions.

\section{Solution of the equations for slow motions}
\label{section-solution}

In what follows, we investigate the initial value problem for the system of
partial differential equations \eqref{maineq}
%\TODO{We need here equation for other variable, not $\theta_n$}
%%% %%% \begin{equation}
%%% %%% \ddot \theta_n+2\eta\dot\theta_n
%%% %%% +\eta^2\theta_n
%%% %%% +
%%% %%% \frac {c^2}4(\theta_{n-1}-2\theta_n+2\theta_{n+1})''=\phi(x,t)\delta_n
%%% %%% \label{maineq}
%%% %%% \end{equation}
%($n\in \mathbb Z$) 
where the heat supply is given in the form of a point source
%\TODO{Do we need this formula?}
\begin{gather}
\chi={\bar\chi}(t)\delta(x),\\
\bar\chi(t)\big|_{t<0}\equiv0,
\label{chi<0}
\end{gather}
supplemented with zero initial conditions stated in the following form
(see 
\eqref{ic<0-pre}):
\begin{equation}
\theta_n(x,t)\big|_{t<0}\equiv0.
\label{ic<0}
\end{equation}

\subsection{The case $\eta=0$}
\label{section-solution-cons}

In this section, we consider a crystal without viscous environment and 
assume that $\eta=0$. First, take
${\bar\chi}(t)=\chil\delta(t)$, where $\chil$ is a constant. 
This corresponds to the choice of heat supply in the form of a point
pulse source. 
Thus,
in accordance with 
Eq.~\eqref{phi-chi},
\begin{equation}
\phi(x,t)=\chil\delta(x)\dot\delta(t).
\label{phi0}
\end{equation}
%For all twice differentiable functions $\varPsi(t)$ the following formulas are
%valid:
%\begin{equation}
%\begin{gathered}
%\partial_t \big(\varPsi(t)H(t)\big)=\varPsi(0)\delta(t)+\dot \varPsi(t)H(t),\\
%\partial_t^2 \big(\varPsi(t)H(t)\big)=\varPsi(0)\dot\delta(t)+\dot\varPsi(0)\delta(t)+\ddot \varPsi(t)H(t).
%\end{gathered}
%\label{0-rule}
%\end{equation}
%Taking here $\varPsi=\theta_n$, one can easily show that system of equations 
%\eqref{maineq}, wherein $\phi$ is defined by 
%\eqref{phi0}, together with initial conditions in form of Eq.~\eqref{ic<0} is 
%equivalent \cite{Vladimirov1971} to initial value problem
%for the homogeneous equation corresponding to Eq.~\eqref{maineq} 
%with following initial conditions:
%\begin{gather}
%\theta_n\big|_{t=0}=\chil\delta(x)\delta_n,\\	
%\dot\theta_n\big|_{t=0}=0.	
%\end{gather}
Now we apply the discrete Fourier transform $\mathscr F_n^\qy$ 
\cite{brigham1974fast,Slepian1980} 
with respect to the variable $n$
to 
%$\theta_n(x,t)$,
Eq.~\eqref{maineq}. This yields
%\TODO{Use something instead $q$}
\begin{equation}
\F{\ddot\theta} \qy n
%+2\eta\dot\theta_\F
-
{\mathscr C^2}\F\theta \qy n''=\chil\delta(x)\dot\delta(t),
\label{1D-wave}
\end{equation}
where
\begin{gather}
\F\theta{\qy}n(\qy,x,t)=\sum_n \theta_n \exp (-\I n\qy),
\label{F-theta-eta0}\\
\mathscr C=c\Big|\sin \frac \qy2\Big|,
\label{CC}
\end{gather}
and $\qy$ is the Fourier transform parameter. 
Here we have used the shift property 
\cite{brigham1974fast,Slepian1980} of the discrete Fourier transform:
\begin{equation}
\F\theta \qy {n\pm1}(\qy,x,t)=\exp(\pm \I\qy)\,\F\theta \qy n(\qy,x,t).
\label{shift}
\end{equation}
Equation \eqref{1D-wave} is the inhomogeneous one-dimensional wave equation.
Therefore, the solution can be written as the convolution of the
right-hand side of \eqref{1D-wave} with the corresponding fundamental
solution \cite{Vladimirov1971}: 
%представляет собой результат применения
%одномерного оператора Д'Аламбера к $\F\theta q n$, поэтому решение
%обобщенной задачи Коши может быть записано как производная по
%времени от соответствующего фундаментального решения \cite{Vladimirov1971}:
\begin{equation}
\F\theta \qy n=
\chil\delta(x)\dot\delta(t)\ast
\frac1{2\mathscr C}\, H(\mathscr Ct-|x|)=
%\frac \chil{2\mathscr C}\,\partial_t H(\mathscr Ct-|x|)=
\frac\chil2 \delta(\mathscr Ct-|x|).
\end{equation}
The inverse of $\F\theta \qy 0$ (the kinetic temperature $T=\theta_0$)
can be expressed in the following form~\cite{brigham1974fast,Slepian1980}:
%Применяя формулу обращения для дискретного преобразования Фурье, получим:
\begin{equation}
\theta_0=\frac 1{2\pi}\int_{-\pi}^\pi \F\theta \qy
n\exp(\I ny)\,dy\,\bigg|_{n=0}=
\frac \chil{4\pi}\int_{-\pi}^{\pi}
\delta\bigg(ct\Big|\sin \frac \qy2\Big|-|x|\bigg)
\,d\qy.
\label{T-trans}
\end{equation}
To calculate the right-hand side of Eq.~\eqref{T-trans}, one needs to use
the formula (see \cite{G-Sh-1})
\begin{equation}
\int_I \delta(f(\qy))\,d\qy=\sum_i \frac{1}{|f'(\qy_i)|},
\label{int-delta}
\end{equation}
where $\qy_i$ are the roots of $f(\qy)$ lying inside the interval $I$. Taking
\begin{equation}
 f(\qy)=ct\Big|\sin \frac \qy2\Big|-|x|,
\end{equation}
one can find the corresponding roots
\begin{equation}
\qy_{1,2}=\pm2 \arcsin\frac{|x|}{ct},\quad ct\geq |x|.
\end{equation}
For $ct<|x|$, there are no roots. 
One has
\begin{equation}
f'(\qy_{1,2})=\frac{ct}2 \cos \frac {\qy_{1,2}}2=\frac 12 \sqrt{c^2t^2-x^2}.
\end{equation}
Applying 
\eqref{int-delta}, one gets
%\TODO{Check this carefully again!}
\begin{equation}
T=\chil\mathfrak F_1(x,t)\=\frac{\chil H(ct-|x|)}{\pi\sqrt{c^2t^2-x^2}}.
\label{T-eta0-0}
\end{equation}
Formula 
\eqref{T-eta0-0} demonstrates that heat propagates at a finite speed $c$.

Now take 
${\bar\chi}(t)=\chio H(t)$, where $\chio$ is a constant.
This corresponds to the choice of heat supply in the form of a
suddenly applied point source of constant intensity.
Thus
\begin{equation}
\phi(x,t)=\chio \delta(x)\delta(t),
\end{equation}
in accordance with 
Eq.~\eqref{phi-chi}. 
%Taking $\varPsi=\theta_n$ in  
%\eqref{0-rule}, one can easily show that system of equations 
%\eqref{maineq}, wherein $\phi$ is defined by 
%\eqref{phi0}, together with initial conditions in form of Eq.~\eqref{ic<0} is 
%equivalent \cite{Vladimirov1971} to initial value problem
%for the homogeneous equation corresponding to Eq.~\eqref{maineq} 
%with the following initial conditions:
%\begin{gather}
%\theta_n\big|_{t=0}=0,\\	
%\dot\theta_n\big|_{t=0}=\chio\delta(x)\delta_n.
%\end{gather}
In this case, an expression for the kinetic temperature can be obtained
by integrating the right-hand side of 
Eq.~\eqref{T-eta0-0} with respect to time:
%\TODO{Discuss asymptotics}
%\TODO{This is not a temperature. Units!}
\begin{equation}
T=\chio\mathfrak F_0\=\frac\chio\pi\int_{|x|/c}^t\frac
{d\tau}{\sqrt{c^2\tau^2-x^2}}=
\frac {\chio\,H(ct-|x|)}{\pi c}\,\ln\frac{ct+\sqrt{c^2t^2-x^2}}{|x|}.
\label{ln-law}
\end{equation}
The formula obtained agrees with previous results \cite{krivtsov2017heat}.
Note that for fixed $x\neq0$, we have $\mathfrak F_0\propto \ln t$ ($\mathfrak F_0$ is
proportional to $\ln t$) as $t\to\infty$.
The solution $\mathfrak F_0$ is
self-similar (it depends only on $x/ct$).

The function $\mathfrak F_0$ is 
{the fundamental solution} for the operator in the left-hand side of 
Eq.~\eqref{maineq} (for $n=0$ and $\eta=0$). %\DANGER
The function $\mathfrak F_1=\dot{\mathfrak F_0}$ plays the role of the fundamental
solution for the problem of the heat propagation in a crystal
without environment caused
by a source of heat supply $\chi(x,t)$.
{The expressions for $\mathfrak F_1$ were earlier obtained in 
\cite{krivtsov2015heat,krivtsov2015unsteady,krivtsov2016ioffe} 
with a slightly
different approach.}
Thus, in the case of an arbitrary function $\chi(x,t)$, the solution $\theta_0$ of 
Eq.~\eqref{phi-chi} that satisfies the zero initial
condition in the form of 
Eq.~\eqref{ic<0} can be written as the convolution 
\begin{multline}
\theta_0=\chi\ast\mathfrak F_1=
\iint_{-\infty}^\infty \chi(\xi,\tau)\,\mathfrak F_1(x-\xi,t-\tau)\,d\xi\,d\tau=\\=
\dot\chi\ast\mathfrak F_0=
\iint_{-\infty}^\infty \dot\chi(\xi,\tau)\,\mathfrak F_0(x-\xi,t-\tau)\,d\xi\,d\tau.
\label{convolution0}
\end{multline}
Here $\ast$ stands for the convolution of functions of two variables $x$ and $t$.
Using formulas 
\eqref{convolution0} in practical applications one should remember that the
time derivative $\dot\chi$ must be calculated in the sense of distributions
(or generalized functions)
\cite{Vladimirov1971}. Also note that the inegration interval in
\eqref{convolution0} is in fact finite due to 
\eqref{chi<0}, 
\eqref{ln-law}.

\subsection{The case $\eta>0$}
\label{section-solution-diss}
In this section, we consider a crystal in a viscous environment and assume
that $\eta>0$. First, take  ${\bar\chi}(t)=\chil\delta(t)$.
This corresponds to the choice of heat supply in the form of a point
pulse source. Thus
\begin{equation}
\phi(x,t)=\phi_0\=\chil\delta(x)\big(\dot\delta(t)+\eta\delta(t)\big),
\label{phi-eta-0}
\end{equation}
in accordance with 
Eq.~\eqref{phi-chi}. 
%Taking $\varPsi=\theta_n$ in  
%\eqref{0-rule}, one can easily show that system of equations 
%\eqref{maineq}, wherein $\phi$ is defined by 
%\eqref{phi0}, together with initial conditions in form of Eq.~\eqref{ic<0} is 
%equivalent \cite{Vladimirov1971} to initial value problem
%for the homogeneous equation corresponding to Eq.~\eqref{maineq} 
%with the following initial conditions:
%\begin{gather}
%\theta_n\big|_{t=0}=\chil\delta(x)\delta_n,\\	
%\dot\theta_n\big|_{t=0}=-\chil\eta\delta(x)\delta_n.	
%\end{gather}
Applying the discrete Fourier transform $\mathscr F_n^y$ 
with respect to the variable $n$
to 
%$\theta_n(x,t)$,
Eq.~\eqref{maineq} and using the shift property 
\eqref{shift}
yields the following equation:
%Применяя к \eqref{maineq} дискретное преобразование Фурье \cite{Slepian1980}
%%%\begin{equation}
%%%\F\theta{y}n(y,x,t)=\sum_n \theta_n \exp (inq),
%%%\end{equation}
%и воспользовавшись формулой сдвига,
%%%\begin{equation}
%%%\F\theta y {n\pm1}(y,x,t)=\exp(\pm iq)\,\F\theta y n(y,x,t)
%%%\end{equation}
%получим
\begin{equation}
\F{\ddot\theta} y n
+2\eta\F{\dot\theta} y n
-
{\mathscr C^2}\F\theta y n''+\eta^2\F\theta yn=
\chil\delta(x)\big(\dot\delta(t)+\eta\delta(t)\big),
\label{1D-telegraph}
\end{equation}
where the symbols
$\F\theta{y}n(y,x,t)$ and 
$\mathscr C$ are defined by Eq.~\eqref{F-theta-eta0} and Eq.~\eqref{CC},
respectively.
The homogeneous equation that corresponds to Eq.~\eqref{1D-telegraph} is
a particular case of the telegraph equation
\begin{equation}
\ddot W+2\eta\dot W-\mathscr C^2W''+\mathscr BW=0.
\label{general-telegraph}
\end{equation}
The fundamental solution for the operator in the left-hand side 
of Eq.~\eqref{general-telegraph} is
(see \cite{polyanin2002handbook})
%\TODO{Introduce $J_0,\ I_0$}
\begin{equation}
\begin{aligned}	
&\Phi=
\frac{e^{-\eta t}H(\mathscr
Ct-|x|)J_0\big(\sqrt{|\alpha|(t^2-x^2/\mathscr C^2)}\big)}{2\mathscr C},&&
\alpha<0,\\
&\Phi=
\frac{e^{-\eta t}H(\mathscr
Ct-|x|)I_0\big(\sqrt{|\alpha|(t^2-x^2/\mathscr C^2)}\big)}{2\mathscr C},&&\alpha>0,
\end{aligned}
\end{equation}
where $\alpha\equiv\eta^2-\mathscr B$. 
The values of the coefficients in
Eq.~\eqref{1D-telegraph}
correspond to the special
limiting case of Eq.~\eqref{general-telegraph} where $\alpha=0$ and 
the fundamental solution is given by the simple formula 
\begin{equation}
\Phi=\frac1{2\mathscr C}\exp(-\eta t)H(\mathscr Ct-|x|).	
\label{f-sol-eta}
\end{equation}
Calculating the convolution of the right-hand side of \eqref{phi-eta-0} with 
the fundamental solution \eqref{f-sol-eta} yields
\begin{equation}
\F\theta yn=\chil(\dot \Phi + \eta \Phi)=\frac\chil2\exp(-\eta t)\delta(\mathscr Ct-|x|),
\end{equation}
therefore,
\begin{multline}
T=\theta_0=
\chil\mathfrak F_1^\eta(x,t)
\=
\frac {\chil\exp(-\eta t)}{4\pi}\int_{-\pi}^{\pi}
\delta\bigg(ct\Big|\sin \frac y2\Big|-|x|\bigg)
\,dy\\=
\frac{\chil H(ct-|x|)\exp(-\eta t)}{\pi\sqrt{c^2t^2-x^2}}.
\label{T-eta-0}
\end{multline}
The function $\mathfrak F_1^\eta$ plays the role of the fundamental
solution for the problem of the heat propagation in a crystal
surrounded by a viscous environment caused by
a source of heat supply $\chi(x,t)$.

Now take 
${\bar\chi}(t)=\chio H(t)$.
This corresponds to the choice of heat supply in the form of a
suddenly applied point source of constant intensity. 
Thus
\begin{equation}
\phi(x,t)=\phi_1\=\chio\delta(x)\big(\delta(t)+\eta H(t)\big),
\label{phi-eta-1}
\end{equation}
in accordance with 
Eq.~\eqref{phi-chi}. 
%\TODO{Show the corresponding IC}
Since $\dot \phi_1=\phi_0$, the non-stationary solution can be obtained
by integrating the right-hand side of 
Eq.~\eqref{T-eta-0}
with respect to time:
\begin{equation}
T(x,t)=
\frac {\chio H(ct-|x|)}\pi \int_{|x|/c}^t \frac{\exp(-\eta \tau)}{\sqrt{c^2\tau^2 -
x^2}}\,d\tau.
\label{result-nonst}
\end{equation}
In contrast to the case $\eta=0$, for $\eta>0$ there exists a stationary
solution which according to 
\cite{PBM1,aleixo2008green}
can be expressed in a closed form:
\begin{equation}
T(x,\infty)=
\frac \chio\pi \int_{|x|/c}^\infty \frac{\exp(-\eta \tau)}{\sqrt{c^2\tau^2 -
x^2}}\,d\tau=\frac\chio{\pi c}\,K_0\left(\frac{\eta |x|}c\right).
\label{basic-result}
\end{equation}
Thus, the stationary spatial profile of the kinetic temperature caused by a
point source of heat supply of constant intensity is described by the
Macdonald function 
(the modified Bessel function of the second
kind)
of zero order.

It may be noted that, using the discrete Fourier transform, the steady-state
solution \eqref{basic-result} can be obtained as the solution $\theta_0(x)$ of
the problem for the static equations
\begin{gather}	
\eta^2\theta_n
+\frac{c^2}4 (\theta_{n-1} - 2\theta_{n} + \theta_{n+1})''
= \eta\chio\delta(x)\de_n
\end{gather}
that correspond to \eqref{maineq} with the boundary conditions at $x\to\infty$
\begin{equation}
\theta_n(x)\to0.
\end{equation}

In the case of an arbitrary function $\chi(x,t)$, the solution $\theta_0$ of 
Eq.~\eqref{phi-chi}
that satisfies the zero initial
condition in the form of 
Eq.~\eqref{ic<0} can be written as the convolution of $\chi$
with the fundamental solution $\mathfrak F_1^\eta$
\eqref{T-eta-0}:
\begin{equation}
\theta_0=\chi\ast\mathfrak F_1^\eta=
\iint_{-\infty}^\infty \chi(\xi,\tau)\,\mathfrak F_1^\eta(x-\xi,t-\tau)\,d\xi\,d\tau.
\label{eta-convolution}
\end{equation}

Thus, we have obtained the analytical solution of the problem.
%% \begin{remark}	
%% \label{remark-fs}
%% Note that if one consider instead of 
%% Eq.~\eqref{phi-chi} an equation of the same structure wherein 
%% $\alpha\neq0$, then 
%% in the case ${\bar\chi}(t)=H(t)$ 
%% the stationary solution 
%% $\theta_0(x,\infty)$ 
%% does not exist, since $T(x,\infty)\propto\ln t$ for
%% fixed $x\neq0$.
%% \end{remark}

\section{Numerics}
\label{Sec-numerics}
In this section, we present the results of the numerical solution of the system of
stochastic differential equations \eqref{1}--\eqref{Winer} 
with initial conditions
\eqref{ic-stochastic}. It is useful to rewrite Eqs.~\eqref{1}--\eqref{Winer}
in the dimensionless form
\begin{equation}
\begin{aligned}	
&d {\tilde{v}}_i =  (\L_{i} {\tilde{u}}_i - \eta {\tilde{v}}_i)
d \tilde t +
\tilde b_i \rho_i \sqrt{d \tilde t},\\ 
&d {\tilde{u}}_i= {\tilde{v}}_i d\tilde t, 
\end{aligned}
\label{dimless}
\end{equation}
where 
\begin{equation}
\tilde u\=\frac u a,\quad \tilde v\=\frac v c,\quad \tilde t\={\omega_0} t,\quad  
\tilde b\=\frac b{c\sqrt{\omega_0}}, \quad \tilde\eta\=\frac\eta{\omega_0}.
\label{all-dless}
\end{equation}
We consider the chain of $2N+1$ particles and the
periodic boundary conditions
\begin{equation}
\begin{aligned}
    u_{-N} &= u_{N-1},  &\qquad &u_{-N+1} = u_N,
    \\
    v_{-N} &= v_{N-1},  &\qquad &v_{-N+1} = v_N.
\end{aligned}
\end{equation}
To obtain a numerical solution in the case of the point source of the heat
supply located at $i=0$, we assume that 
$\tilde b_i\rho_i=\delta_{i0}\tilde b\rho_i$ and
use the scheme 
\begin{equation}
\begin{aligned}	
\Delta {\tilde{v}}_i^j &=  (\L_{i} {\tilde{u}}_i^j - \eta {\tilde{v}}_i^j)
\Delta {\tilde{t}} +
\tilde b\delta_{i0} \rho^j \sqrt{\Delta {\tilde{t}}},\\ 
\Delta {\tilde{u}}_i^j &= {\tilde{v}}_i^{j+1} \Delta {\tilde{t}},
\\
\tilde v^{j+1}_{i} &= \tilde v^{j}_{i} + \Delta \tilde v^{j}_{i}
,
\\
\tilde u^{j+1}_{i} &= \tilde u^{j}_{i} + \Delta \tilde u^{j}_{i},
\end{aligned}
\label{scheme}
\end{equation}
where  $i=\overline{-N, N}$. 
Here the symbols  with
superscript $j$ denote the corresponding quantities at $\tilde
t=\tilde t^j\=j\Delta \tilde t$:
${\tilde{u}}_i^j={\tilde{u}}_i(\tilde t^j),\ \tilde v_i^j=\tilde v_i(\tilde t^j)$;
$\rho^j$ are normal random numbers that satisfy \eqref{82} generated for all
$\tilde t^j$.  Without loss of generality we can take $\tilde b=1$.

We perform a series of $r=1\dots R$ realizations of these calculations (with
various independent 
$\rho^j_{(r)}$) and get the corresponding particle velocities $\tilde v^j_{i(r)}$. In
accordance with
\eqref{temp-def},      
{in order to
obtain the dimensionless kinetic temperature 
\begin{equation}
\tilde T=\frac
{Tk_B}{mc^2},	
\label{kin-temp-dless}
\end{equation}
we should average 
the doubled dimensionless kinetic energies}:
\begin{equation}
\tilde T_i^j =\frac{1}{R}\sum_{r=1}^{R} (\tilde v^j_{i(r)})^2.
\label{kin-temp-num}
\end{equation}

Numerical results  
\eqref{kin-temp-num}
for the kinetic temperature
%obtained for big enough $\tilde t$ 
can be compared with the analytical unsteady solutions
% (in the case $\eta=0$), or
\eqref{result-nonst}, 
\eqref{ln-law}, and steady-state solution
\eqref{basic-result} %(in the case $\eta>0$)
expressed in the dimensionless form:
\begin{gather}
\tilde T({\tilde x},\tilde t)=
\frac{\tilde b^2H(\tilde t-{\tilde x})}{2\pi} \int_{{\tilde x}}^{\tilde t} \frac{\exp(-\tilde\eta \tau)}{\sqrt{\tau^2 -
{\tilde x}^2}}\,d\tau,
\label{result-nonst-dless}
\\
\tilde T({\tilde x},\tilde t)\big|_{\eta=0}
=
\frac {\tilde b^2\,H(\tilde t-{\tilde x})}{2\pi }\,\ln\frac{\tilde t+\sqrt{\tilde
t^2-{\tilde x}^2}}{|{\tilde x}|},
\label{ln-law-dless}\\
\tilde T({\tilde x},\infty)=\frac {\tilde b^2}{2\pi}\, K_0\big(\tilde\eta
|{\tilde x}|\big), 
\label{basic-result-dless}
\end{gather}
where
\begin{equation}
\tilde x \=\frac xa=i.
\label{x-dless}
\end{equation}
Note that the factor $1/2$ in right-hand sides of
Eqs.~\eqref{result-nonst-dless}--\eqref{basic-result-dless}
appears according to Eq.~\eqref{chi-def}.

\begin{figure}[t]	
\centering\includegraphics[width=0.8\columnwidth]{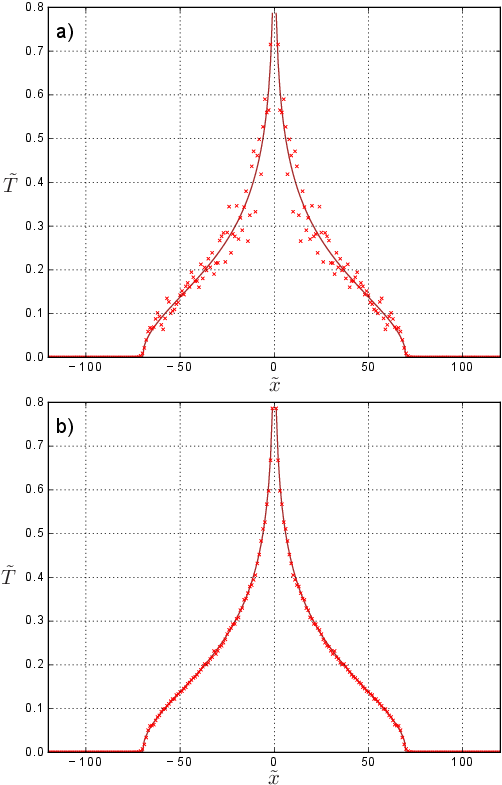}
%%{
%%\psfrag{x}[cb][cb]{$\tilde x$}
%%\psfrag{T}[l][l]{$\tilde T$}
%%}
\caption{Comparing the analytical solution 
\eqref{ln-law-dless} for a crystal without viscous environment (the brown
solid line) and the 
numerical solution (the red crosses) in the case $N=1000$,
$\tilde t=70$. a) $R=100$, b) $R=10000$.}
\label{F12-cons}
\end{figure}

\begin{figure}[t]	
\centering\includegraphics[width=0.8\columnwidth]{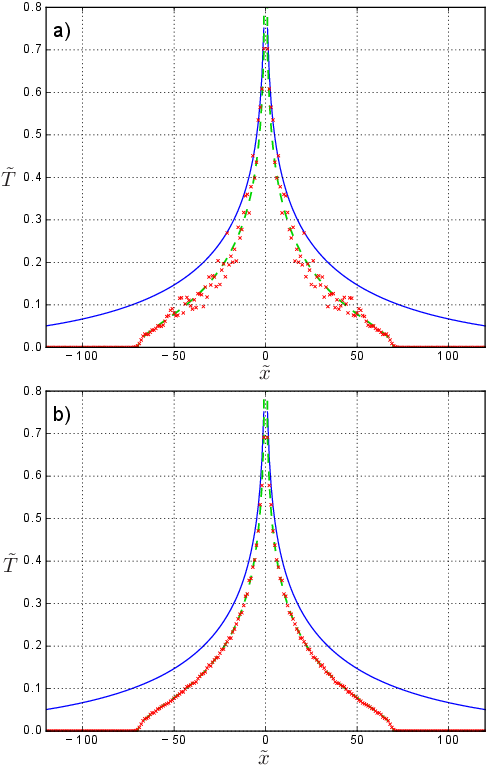}
%%{
%%\psfrag{x}[cb][cb]{$\tilde x$}
%%\psfrag{T}[l][l]{$\tilde T$}
%%}
\caption{Comparing the unsteady analytical solution 
\eqref{result-nonst-dless}
for a crystal in a viscous environment (the green
dashed line), the corresponding steady-state analytical solution 
\eqref{basic-result-dless} in the form of the Macdonald function 
(the blue solid line), and the 
numerical solution (the red crosses) in the case $\tilde\eta=0.01$, $N=1000$,
$\tilde t=70$. a) $R=100$, b) $R=10000$.}
\label{F12-non}
\end{figure}

\begin{figure}[htbp]	
\centering\includegraphics[width=0.8\columnwidth]{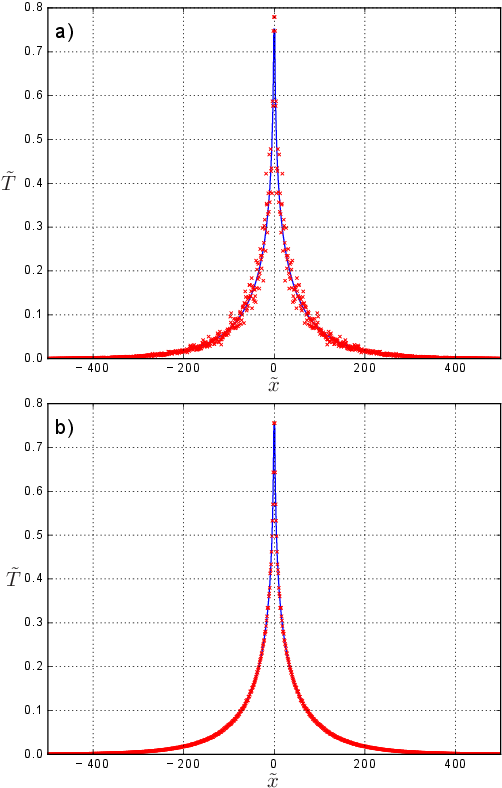}
%%{
%%\psfrag{x}[cb][cb]{$\tilde x$}
%%\psfrag{T}[l][l]{$\tilde T$}
%%}
\caption{Comparing the steady-state analytical solution 
\eqref{basic-result-dless} in the form of the Macdonald function
(the blue solid line) and the
numerical solution (the red crosses) in the case $\tilde\eta=0.01$, $N=1000$,
$\tilde t=500$. a) $R=100$, b) $R=10000$.}
\label{F12}
\end{figure}

%In the case $\eta=0$ one has
%should use formula 
%\eqref{ln-law} in dimensionless form:
%\TODO{Dimensionless heat supply}
%%Here we take into account 
%%\eqref{temp-def},      
%%\eqref{chi-def},
%%and 
A comparison between the analytical and numerical solutions is presented 
in Figures~\ref{F12-cons}--\ref{F12}. 
%\begin{figure}[p]	
%\centering\includegraphics[width=\textwidth]{omega1_0_eta0_0_n1000_T70_0_R100.eps}
%\caption{Comparison between the analytical solution 
%\eqref{ln-law} for the crystal without a viscous environment (the brown
%solid line), and the 
%numerical one (the red crosses) in the case $N=1000$,
%$\tilde t=70$, $\tilde b=1$, $R=100$.}
%\label{F1-cons}
%\end{figure}
%\begin{figure}[p]	
%\centering\includegraphics[width=\textwidth]{omega1_0_eta0_0_n1000_T70_0_R10000.eps}
%\caption{Comparison between the analytical solution \eqref{ln-law} for the
%crystal without a viscous environment (the brown solid line), and the 
%numerical one (the red crosses). The parameters are the same as in
%Fig.~\ref{F1-cons}, but $R=10000$.}
%\label{F2-cons}
%\end{figure}
Figure~\ref{F12-cons} corresponds to a
crystal without viscous environment. 
Figure~\ref{F12-non} correspond to a crystal in a viscous
environment in the case where $\tilde t=70$ is small enough for the
solution to be regarded as an
unsteady one. 
Figure~\ref{F12} correspond to a crystal in a viscous
environment in the case where $\tilde t=500$ is large enough for the solution to be regarded as a
steady-state one.
All figures are presented for two numbers of realizations:
a)~$R=100$ and b) $R=10000$.
One can see that in all cases, for sufficiently large $R=10000$ the analytical and
numerical solutions are in a very good agreement.
%\TODO{Dimensionless quantities on axes}

%\TODO{We need to mark axes on Figures!}
%\begin{figure}[p]	
%\centering\includegraphics[width=\textwidth]{omega1_0_eta0_01_n1000_T70_0_R10000.eps}
%\caption{Comparison between the unsteady analytical solution 
%\eqref{result-nonst}
%for the crystal in a viscous environment (the green
%dashed line), the corresponding steady-state analytical solution 
%\eqref{basic-result}
%(the blue solid line) and the 
%numerical one (the red crosses)
%The parameters are the same as in
%Fig.~\ref{F1-non}, but $R=10000$.}
%\label{F2-non}
%\end{figure}

%
%\begin{figure}[p]	
%\centering\includegraphics[width=\textwidth]{omega1_eta0_01_n1000_T500_R10000.eps}
%\caption{Comparison between the steady-state analytical solution 
%\eqref{basic-result}
%(the blue solid line) and
%numerical one (the red crosses). The parameters are the same as in
%Fig.~\ref{F1}, but $R=10000$.}
%\label{F2}
%\end{figure}

\section{Comparison with the Fourier thermal conductivity}
\label{sec-comparison}
Let us compare our results with the classical results obtained in the
framework of the heat equation based on Fourier's law. 
Consider the case of the non-stationary
temperature distribution caused by a suddenly applied point source of heat
supply. For a crystal in an environment, the solution is given by
formula \eqref{result-nonst}. For large times, there exists a steady-state
solution (see \eqref{basic-result}), in contrast to the case $\eta=0$ of a crystal
without environment, where the solution \eqref{ln-law} of the same
problem grows logarithmically. 
The dimensionless forms of solutions 
\eqref{result-nonst},
\eqref{basic-result},
\eqref{ln-law} 
are 
\eqref{result-nonst-dless},
\eqref{basic-result-dless},
\eqref{ln-law-dless}, respectively. 

Introducing the dimensionless quantities $\tilde t,\ \tilde T, \tilde x$
according to 
\eqref{all-dless},
\eqref{kin-temp-dless}, and 
\eqref{x-dless}, respectively, the classical heat equation in the case under
consideration can be formulated in
the following form 
\begin{equation}
\pd{\tilde T}{\tilde t}-\varkappa \pdd {\tilde T}{\tilde x}=\lambda
H(\tilde t)\delta(\tilde x),
\label{heat-eq-dless}
\end{equation}
where $\varkappa,\ \lambda$ are positive dimensionless constants.
The solution of 
\eqref{heat-eq-dless} that equals zero for $\tilde t<0$ is (see
\cite{Vladimirov1971}) 
\begin{multline}
\tilde T=\frac{\lambda H(\tilde t)}{2\sqrt{\varkappa\pi}} \int_0^{\tilde t}
\frac {\exp\left(-\frac{|\tilde x|^2}{4\varkappa\tau}\right)}{\sqrt \tau}
\,d\tau\\=
\lambda H(\tilde t)\left(
\sqrt{\frac {\tilde t}{\pi\varkappa}}\exp\left(-\frac{\tilde x^2}{4\varkappa
\tilde t}\right)
-\frac{|\tilde x|}{2\varkappa}\erfc\frac{|\tilde x|}{2\sqrt{\mathstrut\varkappa \tilde t}}
\right).
\label{heat-solution}
\end{multline}
For $\tilde t\to\infty$ the right-hand side of
\eqref{heat-solution} grows 
proportionally to $\sqrt {\tilde t}$ being bounded at $\tilde x=0$.

Now we want to compare qualitatively the solutions 
\eqref{result-nonst-dless} 
or 
\eqref{ln-law-dless} 
from the one hand,
and 
\eqref{heat-solution} from the other hand. 
At first, we need to choose the reasonable values for material constants
$\varkappa$ and $\lambda$ in 
\eqref{heat-solution}. In order 
to make the solutions 
corresponding to different physical models more
similar in some sense,
for certain $\tilde t=\tilde t_0$ we
take constants $\varkappa=\varkappa_0(\tilde t_0)$ and
$\lambda=\lambda_0(\tilde t_0)$ such that the following pairs of the quantities 
%\TODO{In fact $\lambda$ does not depend on $t_0$, $\varkappa$ increases with
%$t_0$. I am not sure that we need to discuss this here.}
\begin{equation}
\int_{-\infty}^\infty \tilde T(\tilde x,\tilde t_0)\,d\tilde x
\quad\text{and}\quad
\int_{-\infty}^\infty \tilde x^2 \tilde T(\tilde x,\tilde t_0)\,d\tilde x
\end{equation}
calculated by virtue of 
\eqref{ln-law-dless} and 
\eqref{heat-solution}, respectively, are mutually equal.
In such a way we get that $\lambda_0=1/2$ and does not depend on $\tilde t_0$, while the quantity $\varkappa_0$ depends on $\tilde t_0$.

\begin{figure}[htp]	
\centering{\includegraphics[width=0.8\columnwidth]{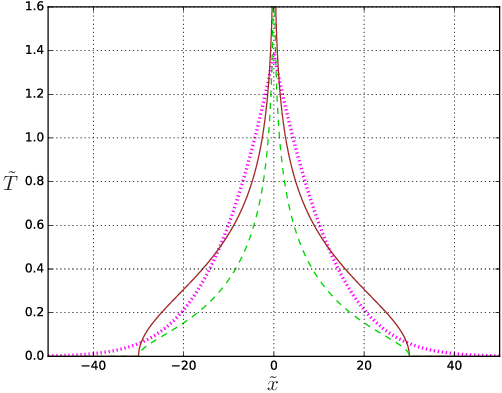}
%%{
%%\psfrag{x}[cb][cb]{$\tilde x$}
%%\psfrag{T}[l][l]{$\tilde T$}
%%}
}
\caption{A qualitative comparison between the non-stationary analytical solution 
\eqref{result-nonst-dless} for a
crystal in a viscous environment ($\tilde \eta=0.03$, the green dashed
line), the non-stationary analytical solution 
\eqref{ln-law-dless}
for a
crystal without viscous environment (the brown solid line), and the solution 
\eqref{heat-solution}
of the
heat equation ($\varkappa=5.00$,
the magenta dotted line).}
\label{F-classic}
\end{figure}

The comparison between unsteady solutions for the crystal and the solution of
the heat equation is given in Figure~{\ref{F-classic}} (all plots are
calculated for $\tilde t_0=30$).
Certainly, the most important difference here is that according to 
\eqref{result-nonst-dless} 
\&
\eqref{ln-law-dless} 
the heat propagates at a finite speed. Both solutions for the crystal are unbounded at
$x=0$, while the solution of the heat equation
\eqref{heat-solution} remains to be finite.

\section{Conclusion}
\label{Sec-conclusion}

In the paper, we started with equations \eqref{1} for stochastic dynamics of
a one-dimensional harmonic crystal in a viscous environment.
We introduced in the standard way the kinetic temperature in the crystal as
a quantity proportional to the statistical dispersion of the particle
velocities. The most important results of the paper are 
the differential-difference equation 
\eqref{maineq}
for the heat propagation in the crystal and
the analytical formulas
\eqref{ln-law},
\eqref{T-eta-0}, and \eqref{basic-result} describing the ballistic heat propagation  in
the crystal from a point heat source.
Formula \eqref{ln-law} gives the non-stationary kinetic temperature
distribution in a crystal without viscous environment caused by a point
source of constant intensity. 
Formulas \eqref{T-eta-0} and \eqref{basic-result} correspond to the case of
a crystal in a viscous environment.
Formula \eqref{T-eta-0} gives the non-stationary kinetic temperature
distribution caused by a point pulse source (i.e., the
fundamental solution).  Formula \eqref{basic-result} shows that
the steady-state
kinetic temperature distribution caused by a point source 
of constant intensity 
is described by 
the Macdonald function of zero
order.
%%\begin{equation*}
%%T=
%%\frac\chio{\pi c}\,K_0\left(\frac{\eta |x|}c\right).
%%\end{equation*}
The comparison between numerical solution of equations \eqref{1} and analytic
solution of differential-difference equation \eqref{maineq} 
demonstrates a good agreement (see Figures~\ref{F12-cons}--\ref{F12}). 
In the case of the heat source of general structure the formula for the
kinetic temperature  
can be obtained as the convolution of the heat source function 
with the corresponding fundamental solution (see Eqs.~\eqref{convolution0},
\eqref{eta-convolution}).

A comparison of our results with the classical model based on the heat equation and
Fourier's law demonstrates an essential difference in the kinetic
temperature distribution near a point source of heat supply
(see Section~\ref{sec-comparison} and Fig.~\ref{F-classic}).
In the framework of our model the heat
propagates at the speed of sound for the crystal.
We expect that the results obtained in the paper can be used to describe the
heat transfer in low-dimensional nanostructures and ultra-pure materials
\cite{chang2008breakdown,xu2014length,goldstein2007mechanics}. On
the other hand, we expect that the theoretical result expressed by formula
\eqref{basic-result} can be verified by experiments with laser
excitation of nanostructures.

\begin{acknowledgements}
The authors are grateful to 
D.A.~Indeitsev, V.A.~Kuzkin, E.V.~Shishkina for useful and stimulating discussions.
\end{acknowledgements}

\appendix

\section{The derivation of the dynamic equations for the covariances}
\label{App-cov}
%Рассмотрим одномерный кристалл, динамика которого описывается уравнением
%\be{1}
%    \dot v_i = F_i + b_i \dot W_i \qq  \dot u_i = v_i,
%\ee
%где $i$ --- индекс частицы, $N$ --- общее число частиц, $u$ и $v$ --- скорость и перемещение, $F_i$ --- сила, действующая на $i$-ую чстицу, $b_i$ --- интенсивность случайного воздействия на соответствующую частицу, $W$ --- винеровский случайный процесс. $F$ и $b$, вообще говоря, могут зависеть от скоростей и перемещений частиц кристалла. Рассматривается как бесконечное число частиц, так и конечное с периодическими граничными условиями.
%Предполагаются периодические граничные условия.
%Начальные условия имеют вид
%\be{2}
%    t=0:\qquad v_i = \si\ro_i \qq  u_i = 0,
%\ee
%где $\ro_i$ --- независимые центрированные случайные числа с единичной дисперсией. Возможна и более общая формулировка с аналогичными начальными условиями для $u_i$.
%
%Дифференциальное уравнение \eq{1} стохастическое, поэтому производные по времени здесь должны пониматься в обобщенном смысле. Альтернативно, система \eq{1} может быть записана для приращений
%\be{3}
%     d v_i = F_i\,dt + b_i \de W_i \qq  d u_i = v_i\,dt,
%\ee
%что несколько более корректно с математической точки зрения. Символ $\de$ при винеровском процессе служит для напоминания, что соответствующее приращение пропорционально не $dt$, а $\sqrt{dt}$, как это требуется для случайных процессов.
%
%\subsection{Уравнения баланса ковариаций}

Consider a system of stochastic differential equations 
\begin{equation}
dx_i = a_i(\mathbf{x},t)\, dt + \sum_\alpha b_{i\alpha}(\mathbf{x},t)\,d
W_\alpha \qq
i=1,...,n,
\label{A4}
\end{equation}
where  $\mathbf{x} = [x_1,\dots,x_n]^\top$ and $W_\alpha$ is a vector of uncorrelated
Wiener variables. 
According to \cite{stepanov2013stochastic} (Chapter 6, formulas (6.4)--(6.17)),
it follows from the It\^o lemma that 
the following equation for the covariance variables holds:
\begin{equation}
    \left\langle x_p x_q\right\rangle^\cdot = \left\langle x_p a_q + x_q a_p + b_{q\alpha} b_{p \alpha}\right\rangle.
\label{A5}
\end{equation}
In the particular case \eqref{1}, we have
$b_{i\alpha}(\mathbf{x},t) = \de_{\alpha i}b_{i}(\mathbf{x},t)$, 
whence
\begin{equation}
    \av{x_p x_q}^\cdot = \av{x_p a_q + x_q a_p} + \de_{pq}\av{b_p b_q}.
    \label{A6}
\end{equation}
Now we apply \eq{A6} to \eq{1} and obtain
\begin{equation}
\begin{gathered}	
    \av{u_p u_q}^\cdot = \av{u_p v_q} + \av{v_p u_q},\\
    \av{u_p v_q}^\cdot = \av{u_p F_q} + \av{v_p v_q},\\
    \av{v_p v_q}^\cdot = \av{v_p F_q} + \av{F_p v_q} + \de_{pq}\av{b_p b_q}.
\end{gathered}
\end{equation}
Applying these formulas to the particular case where
$F_i$ are given by Eq.~\eqref{F_i} yields formula \eqref{5}.

\section{Some identities of the calculus of finite differences}

%                       Additional Definitions for Rnd documents

%-----------------------------------------------------------------------

\def\aav#1{\langle{#1}\rangle}
\def\av#1{\left\langle{#1}\right\rangle}
\def\ax#1{\left\{{#1}\right\}}
\def\AV#1{\left[{#1}\right]}
\def\dv#1{D\!\left({#1}\right)}

%---------------------------------------------------------------------

\let\o=\overline
\let\t=\tilde
\let\h=\hat
\let\c=\check

%---------------------------------------------------------------------

\def\s{^{(s)}}
\let\rro = \varrho

%---------------------------------------------------------------------

\def\D{\Delta}
\def\S{{\cal S}}
\let\f = \varphi
\def\b#1{\stackrel{\smile}{#1}{\!}}
\def\cc#1{\breve #1}

%---------------------------------------------------------------------

\def\Dp{\D\!^+} \def\Dm{\D\!^-}
\def\DD{\D\!^2}
\def\M{{\cal Z}}

%---------------------------------------------------------------------

\def\L{{\cal L}}
\def\figdir{.}
\let\d=\partial

%\rightline{А. М. Кривцов}
%\rightline{07--08 июля 2015 г.}
%\bigskip

%\begin{center}
%\LARGE
%Центральный корреляционный анализ
%\end{center}
%
%\large
%
%\bigskip
%\bigskip
%\centerline{А. М. Кривцов}
%
%\normalsize
%\bigskip
%\centerline{~~07--11 июля 2015 г.}
%
%\vfill
%
%\bigskip
%
%\baselineskip 18pt
%
%\tableofcontents

%\subsection{Обыкновенные операторы}
%
%\subsubsection{Операторы и тождества}

\label{AppB}
Consider an infinite sequence $f_p$, where $p$ is an arbitrary integer.
Introduce the left and right shift operators 
%on a
%%\DANGER
%{half-integer} 
$h$ and $\nu$, respectively:
\begin{equation}
    h f_p \= f_{p+1}
\qq
    \nu f_p \= f_{p-1}
.
\label{AB1}
\end{equation}
%
%Here and in what follows for the sake of simplicity we drop subscript $p$ and
%write just $h$ and $\nu$.
Clearly, 
\begin{equation}
    h \nu = 1
\eqv
    \nu = h^{-1}
.
\label{AB2}
\end{equation}
One has
\begin{gather}
    \L = h + \nu - 2,
    \label{AppB-D2}\\
    \mathscr D=h-\nu.
\end{gather}
%
%
%
%Различные степени оператора сдвига
%\be{3}
%    h^{-1} u_k = u_{k-\frac12}
%\qq
%    h^{0} u_k = u_{k}
%\qq
%    h^{\frac12} u_k = u_{k+\frac14}
%\qq
%    h^2 u_k = u_{k+1}
%.\ee
%Аналогичные формулы выполняются для $\nu$.
By definition, put 
\begin{gather}
    \varSigma \= h + \nu + 2=\L + 4.
%\D \= h - \nu,
%\qquad
%\Sigma \= h + \nu
%%,
%%\qquad
%%\S \= \frac12 \Sigma
%.
%\label{AB3}
\end{gather}
One has
\begin{equation}
\L\varSigma f_p=f_{p+2}-2f_p+f_{p-2}=\mathscr D^2 f_p.
\label{AppBD2}
\end{equation}

%%Разностный, суммирующий и осредняющий операторы
%For the squares of $\D$ and $\Sigma$ one has 
%One has
%\begin{gather}
%    \varSigma = .
%\label{AppB-S2=D2+4}
%%,\\
%%    \S^2 =  \frac14(\D^2 + 4).
%%\label{AppB-SS2=D2+4}
%\end{gather}
%

Introduce the sign change operator $(-1)^p$. One has
\begin{gather}
    h(-1)^p = (-1)^{p+1}h,
\\
    \nu(-1)^p = (-1)^{p+1}\nu,
\\
    \L(-1)^p = (-1)^{p+1}\varSigma,
%\\
%    \Sigma^2(-1)^p = (-1)^{p+1}\D^2,
\\
    \varSigma = (-1)^{p+1}\L(-1)^p.
    \label{AppBD2S2}
%\\
%    \D^2 = (-1)^{k+1}\Sigma^2(-1)^k.
\end{gather}

\selectlanguage{english}

%\bibliographystyle{spmpsci}
%\bibliography{math,thermo,serge-gost}

\end{document}